# Isolated flat bands in an interlocking-circles lattice

Siwen Li[1,2], Yuee Xie[1,2*] and Yuanping Chen[1,2*]

[1]*School of Physics and Electronic Engineering, Jiangsu University, Zhenjiang, 212013, Jiangsu,*

*China*

[2]*School of Physics and Optoelectronics, Xiangtan University, Xiangtan, Hunan, 411105, China*

**Abstract**

Flat-band physics has attracted much attention in recently years because of its interesting properties and important applications. Some typical lattices have been proposed to generate flat bands, such as Kagome and Lieb lattices. The flat bands in these lattices contact with other bands rather than isolated. However, an ideal flat band should be isolated, because isolation is a prerequisite for a number of important applications. Here, we propose a new lattice that can produce isolated flat bands. The lattice is named as interlocking-circles lattice because its pattern seems like interlocking circles. Moreover, the new lattice is realized in graphene by hydrogenation. In the hydrogenated graphene, there are two nontrivial isolated flat bands appearing around the Fermi level. Upon hole or electron doping, the flat bands split into spin-polarized bands and then result in a ferromagnetic graphene. Our work not only proposes a new type of lattice but also opens a way to find systems with isolated flat bands.

---

*Corresponding author: Yuee@xtu.edu.cn and chenyp@ujs.edu.cn.

# 1. Introduction

Flat bands in momentum space have attracted much attention because of amazing physical phenomena behind them[1,2], such as superconducting[3-5], Wigner crystallization[6-8], super solids[9], fractal geometries[10], magnets with dipolar interactions[11], Floquet physics[12] and anomalous quantum Hall effect[13,14]. In a flat band, the carrier' interaction energy dominates over kinetic energy due to the quench of the latter, which is the origination of strong correlation phenomena induce by flat bands[15,16]. Flat bands of ferromagnetism also play a crucial role in determining the behavior of quantum magnets in magnetic fields[17-19].

Typically, a flat band can be induced by localized states[20] or destructive interference[21,22]. The former is common but trivial, which usually appears in the structures with defects and dangling bonds[23,24]. The latter is related to the destructive interference between intrinsic electrons wavefunctions[25], which is nontrivial in most cases[26,27]. Only in some special lattices, such as Kagome[28-30], Lieb[31-33] and coloring-triangle (CT) lattices[34,35], the nontrivial flat bands can be found. An ideal flat band should be isolated, because isolation is a prerequisite for a number of important applications[36,37]. However, the flat bands generated in the lattices mentioned above all contact with other bands. For example, the flat band in a Kagome lattice contact with a quadratic band, while that in a Lieb lattice crosses with a set of Dirac bands[38,39]. Because the contacting or crossing is originated from structural symmetries, the flat bands in the structures are hard to be separated only if the symmetries are eliminated[40]. Therefore, questions arise: can we find a lattice in which there exist intrinsic isolated nontrivial flat bands? if the lattice can be found, can we further find real atomic structures to realize the lattice?

In this paper, we propose a kind of new lattice hosting isolated flat bands. The lattice looks like a pattern of interlocking circles, and thus is named interlocking-circles lattice. Its primitive cell consists of two triangular rings. A tight-binding calculation indicates that there are isolated flat bands in the structure. Moreover, we construct a hydrogenated graphene[41] to realize the interlocking-circles lattice. By a first-principles calculation, two flat bands just appear above and below the fermi level, respectively. Upon hole-doping and electron-doping respectively, the spin degeneracy of the flat bands is spontaneously lifted to result in a flat band ferromagnetism with a spin splitting. Our results will open a gate to find isolated flat bands and new applications in the interlocking-circles lattice.

## 2. Flat bands in an interlocking-circles lattice

The lattice we proposed is shown in Fig. 1(a), which consists of hexagonal rings and triangular rings. The pattern seems like interlocking circles, and thus is named as interlocking-circles lattice. The primitive cell of the lattice is shown in the dashed box in Fig. 1(a), and one can find that the primitive cell is made of two triangular rings.

To describe electronic properties of the interlocking-circles lattice, one can construct a tight-binding (TB) model if each lattice only has one type of electron orbital, say $p_z$. Its Hamiltonian can be expressed as:

$$H_1 = \sum_{i,j} t_{ij} a_i^+ a_j, \tag{1}$$

where $a_i^+$ and $a_j$ represent the creation and annihilation operators of electrons at lattices $i$ and $j$, respectively, and $t_{ij}$ represents the hopping energy between lattices $i$ and $j$. For simplicity, here we only consider the nearest-neighbor hopping energies $t_1$ and $t_2$ [see Fig. 1(a)]. $t_1$ and $t_2$ represent the hopping energies intra- and inter-triangular rings, respectively.

Band structures of the interlocking-circles lattice are dependent on the parameters $t_1$ and $t_2$. When $t_2 \gg t_1$, i.e., the interactions between triangular rings are stronger than those inner the rings, the band structure is shown in Fig. 1(b) ($t_1 = 0.07$ and $t_2 = 0.68$). There are two isolated bands appearing above or below two crossbands around the Fermi level. One can note that, the widths of the flat bands are smaller than the distances between the flat bands and other bands. This satisfies prerequisite for the applications of isolated flat bands[42] . When $t_1 \leq t_2$, no flat bands appear on the band structure or flat bands entangle with other bands [see Fig. S1][43].

## 3. Realization of interlocking-circles lattice on graphene by hydrogenation

One can realize interlocking-circles lattice on graphene by hydrogenation, as shown in Fig. 2. Figure 2(a) and 2(b) exhibit the top and side views of the structure, respectively. The grey atoms are hydrogenated to $sp^3$-hybridization while the blue atoms maintain $sp^2$-hybridization. The primitive cell is shown in the dashed lines in Fig. 2(a), which belongs to a space group of P-3M1. Because the $sp^3$-hybrid atoms have no effect on the electronic properties around the Fermi level, band structure of the hydrogenated graphene is only dependent on the blue atoms. As such, an interlocking-circles lattice is obtained, when two adjacent blue atoms shrink to one lattice (labeled as a yellow ellipse).

The primitive cell contains 18 carbon atoms (blue and gray atoms) and 6 hydrogen atoms (pink atoms) with symmetry constraints of $\alpha = 90°$, $\beta = 90°$ and $\gamma = 120°$. The optimized lattice parameters are $a = b = 7.51$ Å. The hydrogenation not only changes the atomic hybridization from $sp^2$ to $sp^3$, but also results in geometrical distortions. To explain the structure clearly, we have numbered the atoms in Fig. 2(a). In Table 1, the bond lengths between different atoms are given. One can find that the lengths change from 1.365 to 1.532 Å. The shortest bond is that in a yellow ellipse, indicating that the atomic interactions are strongest. The longest bond is that between hydrogenated $sp^3$-hybrid atoms, which is close to the bond length in a diamond. Most carbon atoms are not on the same plane because of distortion. Table 2 presents the fluctuation of atomic locations out of the plane. The largest fluctuation is ±0.43 Å. We also compare the geometric parameters of the hydrogenated graphene[44] with other hydrogenated graphene structures in Table S1[43]. We calculate phonon spectrum of the hydrogenated graphene, as shown in Fig. S2(a) in SI[43]. One can find that no soft phonon mode is found throughout the Brillouin zone (BZ), which indicates that the hydrogenated structure is dynamic stable.

To date, many experiments have been carried out to functionalize graphene[45,46]. Especially, graphene can be hydrogenated to chair configuration and boat configuration[47,48]. Considering the recent fast development of hydrogenated graphene[49,50], the hydrogenated structure we proposed in Fig. 2 could be obtained soon in experiment.

Our calculations were performed within the density-functional theory (DFT) as implemented within the Perdew-Burke-Ernzerhof (PBE) approximation to the exchange-correlation functional[51]. The core-valence interactions were described by the projector augmented wave (PAW) potentials as carried under the Vienna ab initio simulation package (VASP)[52]. The energy cutoff was set to 600 eV. The atomic positions were fully optimized by the conjugate gradient method, and the energy and force convergence criteria were set to be $10^{-6}$ eV and $10^{-3}$ eV/Å, respectively. To avoid interaction between adjacent layers, the vacuum distance normal to the layers was kept to 20 Å. Integrations over the Brillouin zone were done with a 5×5×1 Γ-centered Monkhorst-Pack k-point mesh. The phonon calculations were carried out using the Phonopy package with the forces calculated by the VASP code.

To evaluate the thermal stability, we carried out ab initio molecular dynamics (AIMD) simulations with canonical ensemble, for which a 2 × 2 supercell containing 96 atoms was used and the AIMD simulations were performed with a Nose-Hoover thermostat at 800 and 900 K, respectively. After heating up to the targeted temperature 800 K for 20 *ps*, we did not observe any structural decomposition (Fig. S2(b)) [43]. The structure reconstruction only occurs at 900 K during the 20 *ps* simulation. Therefore, our calculations from phonon dispersion and AIMD simulations fully indicate that the hydrogenated graphene has rather high thermodynamic stability, and outstanding thermal stability.

## 4. Flat bands and magnetism in hydrogenated graphene

Figure 3 shows the electronic energy band structure and density of states of the hydrogenated graphene in Fig. 2. There are two flat bands on the band structure: one is just below the Fermi level, the other appears at 1.66 eV. Meanwhile, two sets of crossing bands locate at two sides of the two flat bands, respectively [see Fig. 3(a)], and the degenerate crossing points are protected by the crystal symmetries. The partial density of states (PDOS) in Fig. 3(b) indicates that the flat bands and crossing bands are mainly attributed by $p_z$ orbitals. Figure 3(c) present the $p_z$ orbitals distributions of the PDOS on different atoms. One can find that the energy bands around the Fermi level are mainly induced by the atoms 1 ~ 12, i.e., *sp²*-hybrid atoms. The charge densities of two quantum states on the two flat bands further illustrate that the electrons on the flat bands are localized on the *sp²*-hybrid atoms. The *sp³*-hybrid atoms look like a circle wall that interrupts the interactions between the localized states.

To explain the origination of the flat bands, we simulate the electronic properties of the hydrogenated graphene in Fig. 2 by a tight-binding model like Eq. (1). As mentioned above, because the electronic properties are mainly determined by the $p_z$ orbitals on the $sp^2$-hybrid atoms, one can construct a tight-binding Hamiltonian based on the model in Fig. 4(a):

$$H_2 = \sum_i \varepsilon_i + \sum_{i,j} t'_{ij} a_i^+ a_j, \qquad (2)$$

where $\varepsilon_i$ represents the on-site energy at lattice $i$, and $t'_{ij}$ represents the hopping energy between lattices $i$ and $j$. Here, we only consider four hopping energies: $t'_1$, $t'_2$, and $t'_3$ describing the nearest-neighbor interactions, and $t'_4$ describing a next-nearest-neighbor interaction, as shown in Fig. 4(a).

Figure 4(b) shows a band structure of Eq. (2) with the parameters $t'_1 = 2.07$, $t'_2 = 1.76$, $t'_3 = 0.1$, $t'_4 = 0.08$ and $\varepsilon_i = 0.9$ ($i = 1 \sim 12$). One can find that the energy bands in Fig. 4(b) are equal to two sets of energy bands in Fig. 1(b): one set is above the Fermi level and the other set is below. This is because each interlocking-circles lattice in Fig. 4(a) is made of two atoms. The strong interactions $t'_1$ results in the separation of the two sets of bands. This demonstrates that the hydrogenated graphene is a good example of the interlocking-circles lattice. Comparing the projected energy bands of $p_z$ orbitals calculated by DFT method in Fig. 4(c), one can find that the tight-binding results in Fig. 4(b) fit well with the DFT results.

Based on Stoner criterion[53], the ferromagnetism can arise from some localized states under partial filling. Figure 5 shows band structures of the hydrogenated graphene under a hole or electron doping. Figure 5(a) is the case of a hole doping, i.e., the flat band below the Fermi level is half filling. One can find that the spin-up and spin-down energy bands split (about 0.15 eV) and the Fermi level crosses the flat bands. The hydrogenated graphene changes to a ferromagnetic state with a direct exchange mechanism. Our calculations indicate that the total magnetic moment is 0.44 $\mu_B$. The band structure of one electron doping is given in Fig. 5(b). In this case, the flat band above the Fermi level is half filling. We can also see a significant spin splitting of 0.15 eV, and the total magnetic moment is also 0.44 $\mu_B$. In the experiments, electron or hole doping can be achieved by electrostatic gating[54].

## 5. Conclusions

In summary, we propose a new lattice named as interlocking-circles lattice, which can produce nontrivial isolated flat bands. The lattice is made of triangular rings. A tight-binding model is used to exhibit electronic properties of the lattice. There are two isolated flat bands in the band structure. Moreover, we construct a hydrogenated graphene to realize the interlocking-circles lattice. By a first-principles calculation, two isolated flat bands appear just around the Fermi level. Upon hole or electron doping, the spin degeneracy of the flat bands is spontaneously lifted and then results in a ferromagnetic graphene. Considering the recent fast development of hydrogenated graphene[55-57], the hydrogenated graphene could be obtained soon in experiment.

Our work opens a way to find isolated flat bands by the interlocking-circles lattice. Here, we only use hydrogenated graphene as an example to realize interlocking-circles lattice. We believe that the new lattice can be also realized in some other 2D materials, and we also hope some experiments can be carried out to try to synthesize the new type of materials.

## Acknowledgments

This work was supported by the National Natural Science Foundation of China (No. 12074150, No. 11874314).

**Figure Captions**

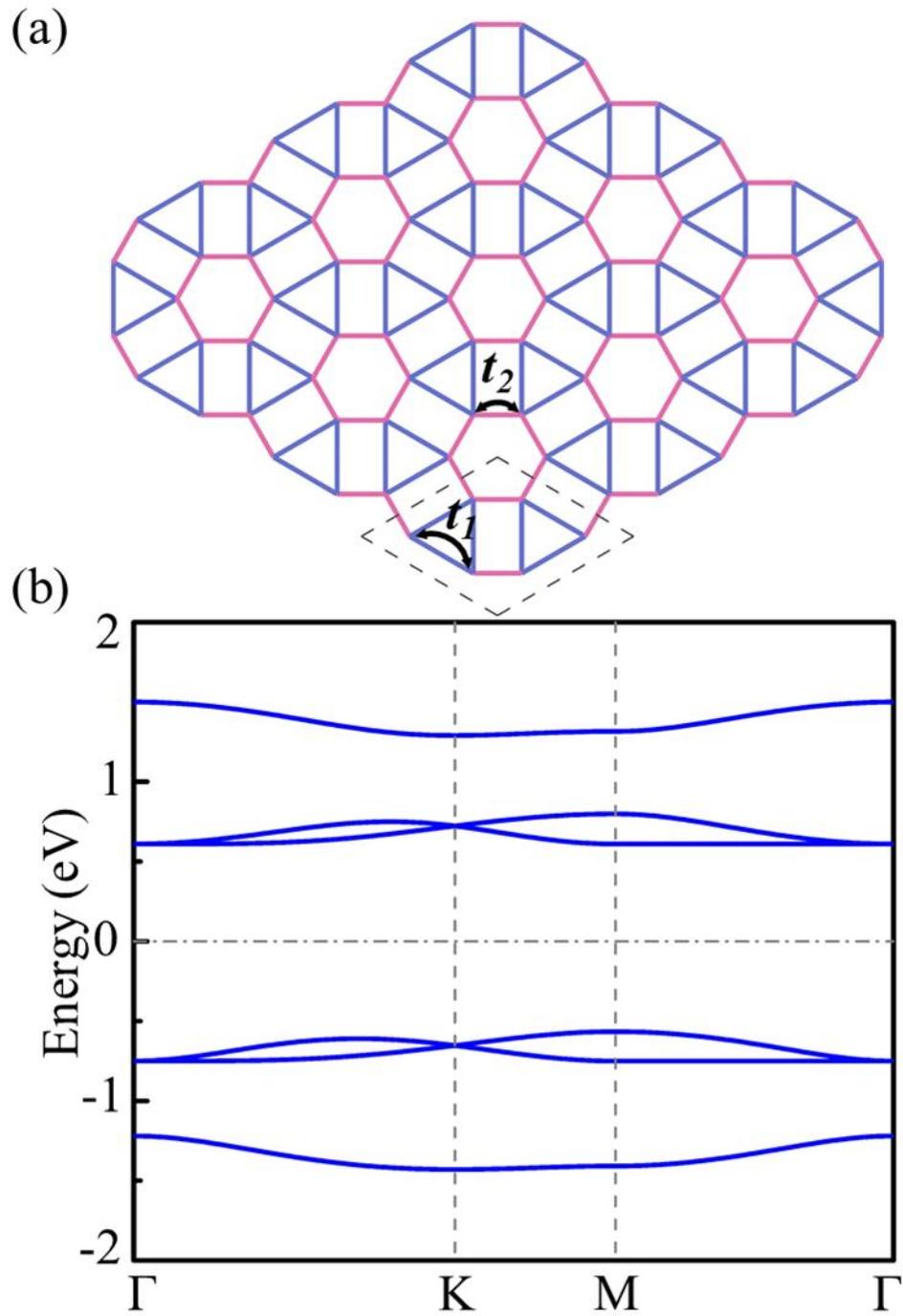

**FIG.1.** (a) An interlocking-circles lattice, where the primitive unit cell is shown by the dashed line. $t_1$ and $t_2$ represent the hopping energies intra- and inter-triangular rings, respectively. (b) Band structure based on the tight-binding model in Eq. (1) with $t_1 = 0.07$ and $t_2 = 0.68$.

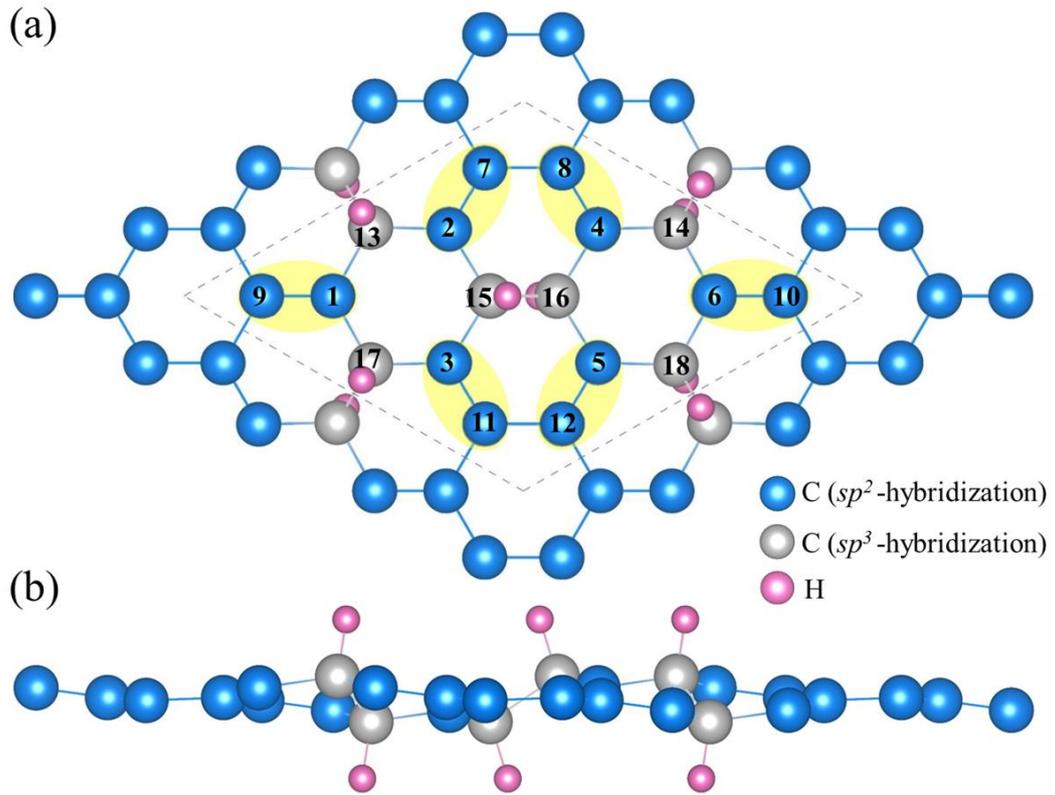

**FIG.2.** Top view (a) and side view (b) of hydrogenated graphene, where the blue and gray dots represent $sp^2$- and $sp^3$-hybrid C atoms while the pink dots represent H atoms. Its primitive cell is shown by the dashed lines in (a).

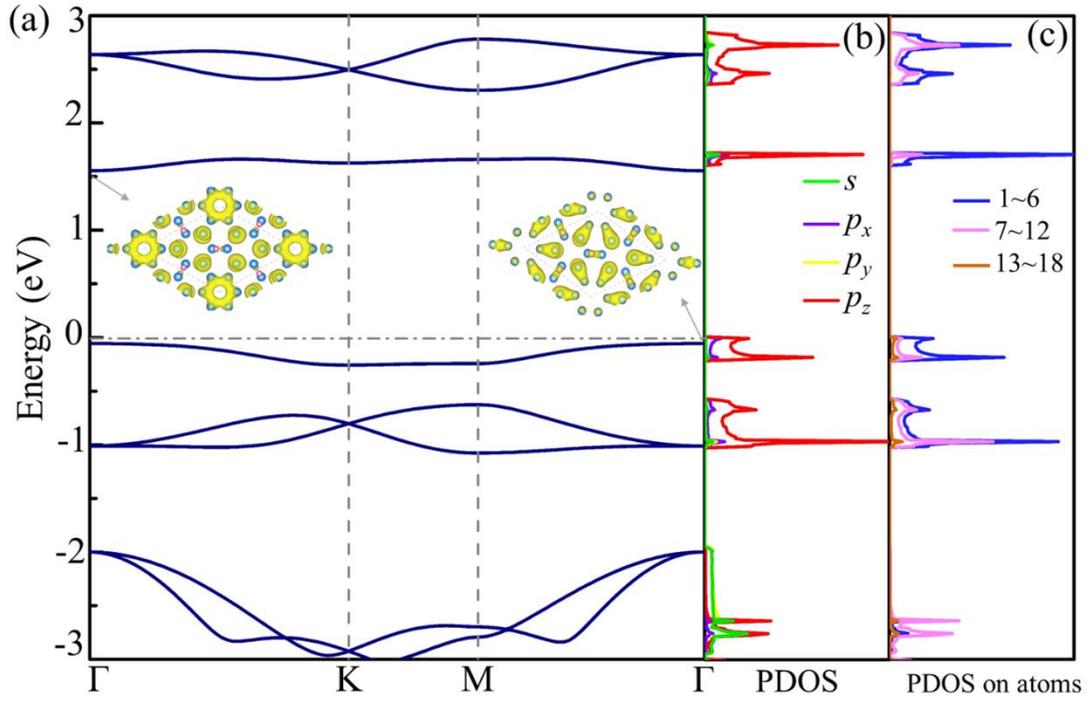

**FIG. 3.** (a) Band structure of the hydrogenated graphene in Fig. 2. Insets: charge densities for the quantum states on the flat bands. (b) PDOS of the hydrogenated graphene for the different orbitals. (c) Distributions of the PDOS for the pz orbitals on different atoms.

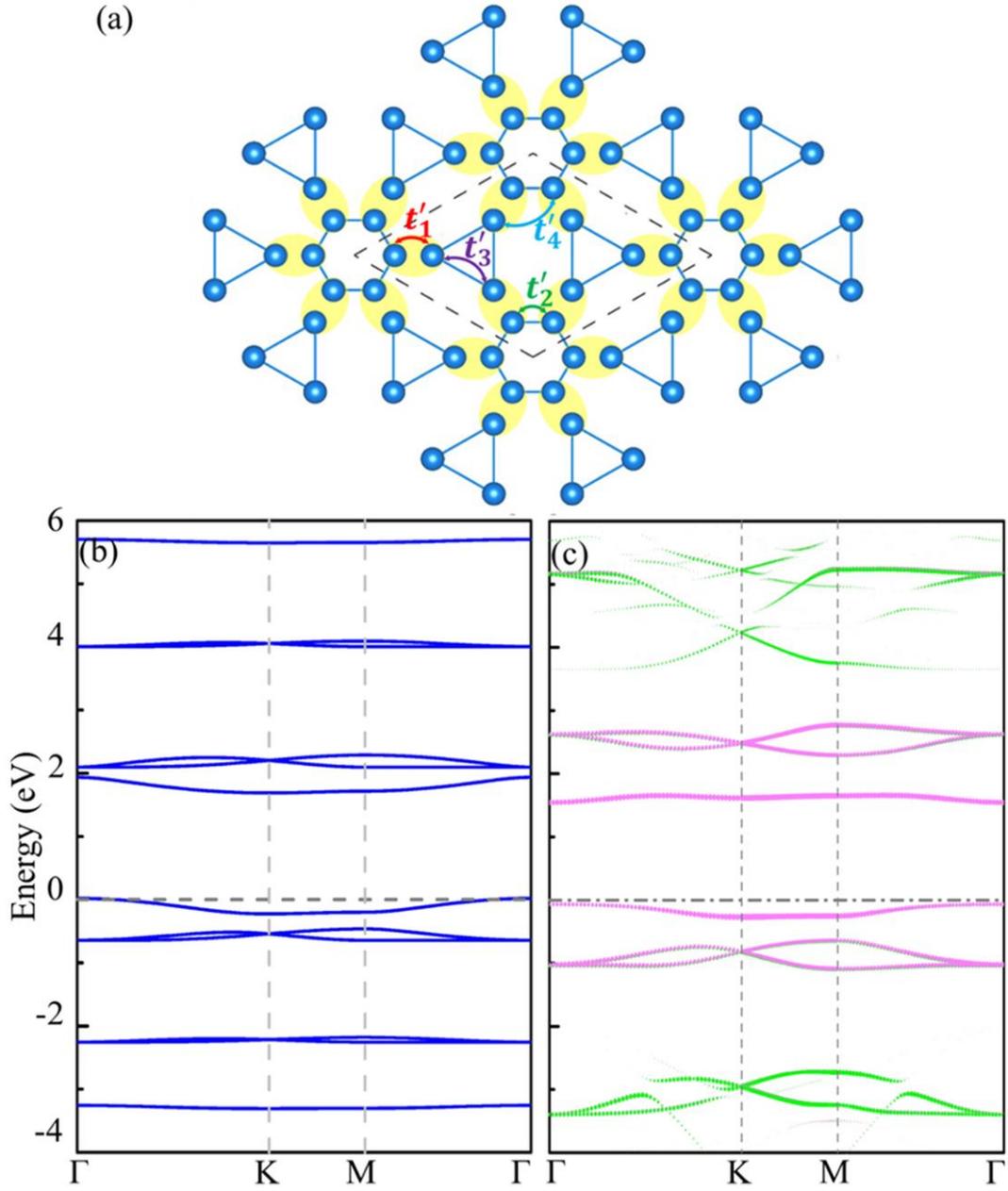

**FIG. 4.** (a) A simplified model for the hydrogenated graphene in Fig. 2, where only $sp^2$-hybrid carbon atoms are retained. $t'_1$, $t'_2$, and $t'_3$ describe the nearest-neighbor interactions, and $t'_4$ describes a next-nearest-neighbor interaction. (b) Band structure based on the tight-binding model in Eq. (2). (c) Projected energy bands for $p_z$ orbitals calculated by DFT method.

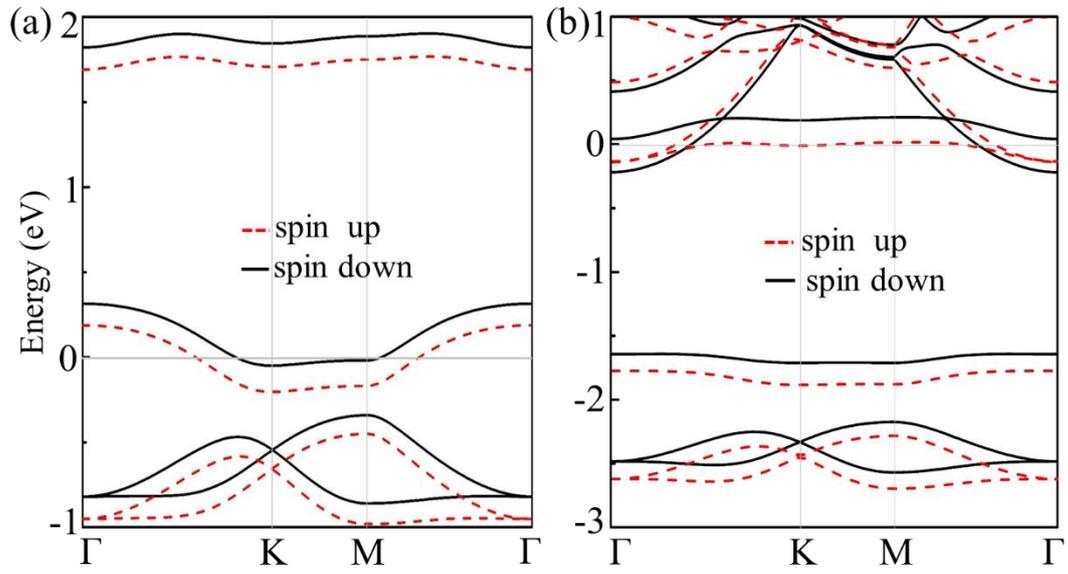

**FIG. 5.** (a) and (b) Bands structures of the hydrogenated graphene in Fig. 2 after a hole doping (a) or an electron doping (b). The dashed and solid lines represent spin-up and spin-down energy bands, respectively.

Table 1. The bond lengths between different atoms of hydrogenated graphene in Fig. 2.

| atomic number | 1-9 | 1-13 | 1-17 | 2-7 | 2-13 | 2-15 | 3-11 |
|---|---|---|---|---|---|---|---|
| Bond Lengths (Å) | 1.365 | 1.503 | 1.503 | 1.365 | 1.503 | 1.503 | 1.365 |
| atomic number | 3-15 | 3-17 | 4-8 | 4-14 | 4-16 | 5-12 | 5-16 |
| Bond Lengths (Å) | 1.503 | 1.503 | 1.365 | 1.503 | 1.503 | 1.365 | 1.503 |
| atomic number | 5-18 | 6-10 | 6-14 | 6-18 | 7-8 | 15-16 | 11-12 |
| Bond Lengths (Å) | 1.503 | 1.365 | 1.503 | 1.503 | 1.467 | 1.532 | 1.467 |

Table 2. The fluctuation of atomic locations out of the plane in Fig. 2. Atoms 7-12 locate on the plane. The upward fluctuation is labeled as "+", while the downward fluctuation is labeled as "-".

| atomic number | 1 | 2 | 3 | 4 | 5 | 6 | 7 | 8 | 9 |
|---|---|---|---|---|---|---|---|---|---|
| Undulating distance (Å) | 0.16 | 0.20 | 0.20 | -0.20 | -0.20 | -0.16 | 0 | 0 | 0 |
| atomic number | 10 | 11 | 12 | 13 | 14 | 15 | 16 | 17 | 18 |
| Undulating distance (Å) | 0 | 0 | 0 | 0.41 | -0.41 | 0.43 | -0.43 | 0.41 | -0.41 |